\documentstyle[prb,aps]{revtex}
\begin{document}
\input{psfig} 

\title{Learning effective amino acid interactions through iterative
stochastic techniques} 

\author{Cristian Micheletti$^1$, Flavio Seno$^2$, Jayanth
R. Banavar$^3$ and Amos Maritan$^1$}
\vskip 0.3cm
\address{(1) International School for Advanced Studies and INFM,
Via Beirut 2, 34014 Trieste, Italy}
\address{(2) INFM -Dipartimento ``G. Galilei'' , Via Marzolo, 8
Padova, Italy}
\address{(3) 104 Davey Laboratory, The Pennsylvania State University,
University Park, Pennsylvania 16802}
\date{\today}
\maketitle

\tighten
\begin{abstract}
 The prediction of the
three-dimensional structures of the native state of proteins from the
sequences of their amino acids is one of the most important
challenges in molecular biology. An essential ingredient to solve this
problem within coarse-grained models is the task of deducing effective
interaction potentials between the amino acids. Over the years several
techniques have been developed to extract potentials that are able to
discriminate satisfactorily between the native and non-native folds of
a pre-assigned protein sequence. In general, when these potentials are
used in actual dynamical folding simulations, they lead to a drift of
the native structure outside the quasi-native basin. In this study, we
present and validate an approach to overcome this difficulty. By
exploiting several numerical and analytical tools we set up a rigorous
iterative scheme to extract potentials satisfying a pre-requisite of
any viable potential: the stabilization of proteins within their
native basin (less than 3-4 \AA$\ $ cRMS). The scheme is flexible and is
demonstrated to be applicable to a variety of parametrizations of the
energy function and provides, in each case, the optimal potentials.
\end{abstract}

\vskip 2.0cm

\section{Introduction}

\subsection{Background}

In recent years, there have been numerous studies pertaining to the
determination of effective amino acid interactions for coarse-grained
models of proteins
\cite{Crippen91,Crippen96,Fri89,Maiorov92,Dima99,Miyazawa85,Miyazawa96,Mirny96,Moult96,Seno98a,Sippl90,Sippl95,Tanaka76,Thomas96a,Thomas96b,Karplus2000}.
The reasons are two-fold. If reliable interaction potentials were
available, they could be used, in principle, to predict the native
state of a known protein sequence through energy minimization
techniques -- a possibility underscored by the experimental
observation of Anfinsen \cite{Anfinsen73} that proteins fold rapidly
and reversibly into a unique conformation which he postulated to be
the free energy minimum. Second, the knowledge of good potentials is a
necessary ingredient to perform design of protein-like sequences,
i.e. finding the sequence which has the assigned target structure as
its native state
\cite{Bowie91,Deutsch96,Giugliarelli2000,Micheletti98a,Micheletti98b,Micheletti99a,Morrissey96,Pabo83,Quinn94,Seno96,Seno98b,mayo}.

A simple and popular method for the extraction of the interaction
potentials is the so-called quasi-chemical approximation
\cite{Miyazawa85,Miyazawa96,Sippl90,Sippl95}, which infers the
strength of pairwise interactions from the relative abundance of
distinct pairs of amino acids in contact. In favour of this technique
are its simplicity of implementation and robustness against use of
different sets of proteins used to extract the data. The quasichemical
approach is, however, an approximate scheme since it neglects the
peptide bonding of the amino acids by treating them as a gas. In this
respect it can be viewed as a first correction to a sort of mean-field
approximation
\cite{Seno98a,Thomas96a,Thomas96b,du,Jort99}. Quasi-chemical methods
have proved valuable since they can score satisfactorily as far as
thermodynamic stability is concerned. In fact, they usually assign low
energies to the native state of a sequence compared to the mean energy
of the same sequence mounted on unrelated structures of the same
length.

Other alternative extraction strategies
\cite{Crippen91,Crippen96,Dima99,Seno98a,Jort99} pioneered by Crippen
\cite{Crippen91} use the thermodynamic stability criterion as the
extraction method itself rather than as a mere validating tool. These
schemes aim at finding a set of potentials so that given a protein
sequence, its native state is recognized as having an energy
well-below other conformations of the same length (decoys). A major
advantage of such schemes is the possibility \cite{Jort99} of
verifying directly whether the chosen parametrization of the free
energy is appropriate. In fact, if the energy function introduced to
describe the system is too simplistic, then it will not be possible to
adjust its parameters so that the native states of all the proteins
have lower energy than each of the competing decoy conformations.  In
fact, recent studies \cite{Vendruscolo98} have pointed out that a
pairwise energy function may not stabilize even a single protein such
as crambin.  There are at least two explanations for the failure to
learn good sets of potentials: one possibility is that structures too
similar to the native one have been included in the training set of
decoy conformations. The other, more serious concern, is that the
parametrization for the energy function is too simple to capture the
physics of the problem. Therefore, a given set of proteins might be
unlearnable with a particular energy function (and strictly speaking
might remain unlearnable with any energy function if infinite
precision is required).

A key difficulty in implementing this powerful procedure is in the
careful choice/generation of the decoy structures. In many instances,
the decoys are generated by taking compact ``chunks'' of suitable
length from a bank of proteins (threading) \cite{threading}.  Such
decoys may not be physical for certain sequences because of steric
constraints and are usually not very stringent, i.e. they do not
compete significantly with the native state to be occupied below the
folding transition temperature. This ends up with placing rather loose
or unphysical constraints on the extracted potentials.

Both the potentials extracted with the latter approach or the
quasi-chemical one, undoubtedly capture the main features of amino
acid chemistry \cite{necprl} and of the folding process \cite{du}.
However, they have a range of applicability limited within the same
scheme that was used for extraction. For example, the performance of
the potentials determined using the quasi-chemical method is
unsatisfactory when they are applied to unbiased folding simulations.
In particular, dynamical trajectories starting from a protein's native
state well below the folding transition temperature always escape from
the quasi-native basin (with a 3-4 \AA\ root mean square deviation per
site \cite{huang}).

An alternative strategy \cite{Clementi98a} would be to start from a
trial set of potentials, carry out repeated folding processes with the
aid of a computer and find the structures with the lowest energy. If
any of these structures are significantly different from the native
ones, then the potentials are modified so that their energy is
increased above the target native state (destabilization). This
process can be repeated until the native state is fully stabilized.
More generally, the process should be carried out simultaneously for a
set of non-homologous proteins.  A key difficulty of the strategy is
the need to have an efficient folding algorithm, which, in principle,
could lead to the folding of a given sequence from the denatured state
into its native state conformation corresponding to a given set of
interaction parameters.  In other words, if one has a powerful folding
alogorithm, one can tune the parameters of the potential just right to
ensure that the native states of a set of proteins are accurately
learnt.

\subsection{A new iterative strategy}

Here, we will present a general strategy for determining the effective
interactions between the coarse-grained degrees of freedom of a
protein.  Our procedure does not entail the difficulties associated
with the methods described above.  The key idea is the observation
that the native states of proteins must at least satisfy the
Anfisenian requirement of being located at the bottom of a smooth free
energy minimum \cite{Bry,Woly} with a wide basin of attraction
\cite{funnel2} This suggests a straightforward approach:

\begin{enumerate}
\item Begin with an initial guess of the potential parameters.
\item Start from the native states of several proteins and carry out
an unbiased Monte Carlo (or molecular dynamics) simulation (say at
zero temperature) and determine several accessible local minima for
each of the proteins.
\item Modify the potential parameters in such a way as to destabilize
these conformations in favor of the known native state conformations.
\item Iterate this procedure by returning to (2).
\end{enumerate}

After several iterations, one would expect to converge to a set of
potential parameters which best capture the optimal shape of the free
energy landscape in the vicinity of all the native state structures.

Our method adopts the thermodynamic stability scheme described before
but with the proviso that the decoys would be generated by an explicit
and simple dynamical process. The structures so generated are
guaranteed to be stringent competitors of the native structure in
housing the sequence below the folding transition temperature.

Our strategy is a basic pre-requisite for viable effective
potentials \cite{huang,god} and hence is an obligatory step along the
difficult route to fully-automated folding prediction.  Since the
scheme is both flexible and optimal, it can be used to compare the
performance of many different scoring functions or parametrizations
and hence select the most
promising one for ab initio folding simulations.

A fundamental question pertaining to our new strategy is ``How well
can the native state be stabilized for a given form of the energy
function?''. In this article, we will tackle this question and
determine the optimal choice of the parameters for a given energy
function in order to approach the native states as well as possible.

We proceed by considering a set of 20 single chain proteins (training
set) taken from different protein families and introduce a pairwise
energy function to describe the interactions between the amino acids.
Then, by using an iterative procedure, we systematically modify the
parameters characterizing the pairwise potentials to optimize the
local stability of the native states. It is shown that the iterative
procedure brings a systematic improvement of the quality of the
potentials.  The optimal potentials confer significant thermal
stability .  Moreover we check the quality of our potentials by
testing them on sets of hundred of decoys believed to be very
stringent, obtained by Levitt for seven heavily investigated
proteins \cite{Park95}.  The results of the tests are found to be very
encouraging.

The main message of our paper is to present and demonstrate the
viability of the idea that the effective potential between amino acids
can be learned by ensuring the local stability of the native states of
many proteins simultaneously.  There are obviously many ways of
implementing this idea and we will present a few schemes here that we
have tried and which yield remarkably good results.

\section{Theory}

\subsection{The model}

Microscopic molecular dynamics techniques using presently available
computational resources can follow the dynamics of short peptides for
time scales significantly shorter than typical folding times. For
these reasons, it has become customary to simplify both the
representation of protein conformations and their dynamical behaviour
\cite{Micheletti99b}.  Most of the procedures adopted to coarse grain
the microscopic degrees of freedom of proteins substitute a whole
amino acid with an effective centroid placed at a suitable point along
the CA-CB direction, or coinciding with one of these two atoms. This
choice, that we shall also adopt in this study brings about a drastic
simplification of the structure of the energy function involving only
interactions between the centroids. The coarse-graining procedure that
we adopted is inspired by the XFCC model of Covell and Jernigan
\cite{Covell}. According to this model, both the dihedral angles and
peptide-bond length are discretized so that there are several degrees
of freedom per amino acid (identified with CA atoms). This framework
allows a faithful representation of protein backbones, since the
coarse grained CA positions are typically within 1 \AA $\ $ of the
crystallographic positions. The preservation of the typical angles is
carried out at the expense of variations in the peptide bond length,
$d$, (the separation between consecutive CAs) which can be stretched
within the bounds $2.6 \AA < d < 4.7 \AA$.  Within such a range of
$d$, there are up to 40 possibilities for placing a CA centroid. This
makes the Covell-Jernigan model probably the best compromise between
having a description as close as possible to the continuum while
retaining only few degrees of freedom \cite{Park95}.  Besides the
CA's, we also introduce CB atoms constructed using a geometric rule
(inspired from one obtained in the continuum \cite{Park96}) deduced
from peptide geometry in the XFCC context:

\begin{equation}
\vec{r}^{CB}_{i} = l(\hat{a} \cdot \cos{\theta} + \hat{b} \cdot \sin{\theta})
\end{equation}

\noindent where:

\begin{equation}
\hat{a}= \frac{\hat{s}_{i,i-1} + \hat{s}_{i,i+1}}{|\hat{s}_{i,i-1}+
\hat{s}_{i,i+1}|} \ \ \ \ \ \
\hat{b}= \frac{\hat{s}_{i,i-1} \wedge {s}_{i,i+1}}{|\hat{s}_{i,i-1} \wedge
\hat{s}_{i,i+1}|}  .
\end{equation}

In the previous equations, $\hat{s}_{i,j}$ is the unit vector:

\begin{equation}
\hat{s}_{i,j}= \frac{\vec{r_i}^{CA}-\vec{r_j}^{CA}}{|\vec{r_i}^{CA}
-\vec{r_j}^{CA}|}
\end{equation}

\noindent while $l$ is the distance of the CB atom from the CA atom
that we choose equal to 3 \AA, $\theta$ is an appropriate angle
optimally chosen to be $37.6^0$ and $\vec{r}^{CA(CB)}_k$ is the
position of the $k$--th CA(CB) atom along the chain. When
crystallographic positions are used, such restrictions place the CB
within 0.3 \AA\ of the true position; when discretized values for CA
locations are used, the discrepancy is increased to about 1 \AA.

\subsection{Parametrization of potential energies of interaction}

Most of the results, that we present here, are for a simple 
energy function with interactions only
between pairs of non-consecutive CA atoms,

\begin{equation}
{\cal H} = \sum \Delta (| \vec{r}_i^{CA} - \vec{r}_j^{CA}|)
\cdot \epsilon(S_i, S_j) + 10 \cdot \epsilon_r \cdot
\left[ \left( \frac{4.65}{(| \vec{r}_i^{CA} -
\vec{r}_j^{CA}|)} \right)^2 -1 \right] \cdot \Theta(4.65 - | \vec{r}_i^{CA} - \vec{r}_j^{CA}|)
\end{equation}
\noindent where, 

\begin{equation}
\Delta (r) = \frac{1}{2} + \frac{1}{2} \tanh{\frac{6.5-r}{2}  }
\end{equation}

\noindent and $\Theta$ is the step function. We will refer to this
model as the model A.

$\Delta$ denotes the distance-dependent strength of interactions
between the $i$-th and the $j$-th amino acids along the sequence
mounted on the structure $\Gamma$, $\epsilon$ is the interaction
matrix and $S_i$ denotes the type of the $i$th amino acid in the
sequence.  $\epsilon_r$ is a repulsive term that penalizes cases where
two non-consecutive pairs of CA's are closer than 4.65 \AA, a
circumstance rarely encountered in protein structures. Altogether,
there are 211 parameters to be learnt (with all of them eventaully
scaled by the normalization condition).

\subsection{Monte Carlo dynamics}

The dynamics in conformation space is carried out using a Monte Carlo
technique. A newly generated conformation is accepted according to the
standard Metropolis rule. At each attempted MC step, we move up to 2
of consecutive protein residues to unoccupied discrete positions. The
new positions are constrained to satisfy a set of suitable physical
constraints that we deduced by a statistical analysis of the CA and CB
positions of an ensemble of over hundred single-chain globular
proteins within the XFCC model. More precisely:

\begin{itemize}
\item The separation $d$ between two consecutive $CA$ atoms (measured
in \AA) must remain in the range $ 2.6 < d < 4.7 $;

\item Two non-consecutive $CB$ atoms must not be closer than 2\ \AA.

\item Two non-consecutive $CA$ and $CB$ atoms must not be closer than
2 \AA.

\item The chain length is allowed to fluctuate by up to a maximum value of
4 \AA (in magnitude) with respect to the original length.
\end{itemize}

The Monte Carlo algorithm was used to relax the crystallographic
structure (XFCC-discretized) to its lowest energy states. This is
conveniently done by setting the MC temperature close to zero and
carrying out typically 300,000 attempted moves.  The relaxed
configurations are used as decoy structures for refining the
potentials by requiring that the native state has lower energy than
the relaxed ones. This amounts to a requirement that the optimal
potentials should, at least, ensure the best local stability for the
protein.

\subsection{Finding the optimal potentials}

A convenient way to find the optimal potentials is the use of the
perceptron algorithm \cite{perceptron} for the optimization
of a set of linear inequalities.

In our case, the inequalities are of the form

\begin{equation}
{\cal H}(S,\Gamma) - {\cal H}(S,\Gamma_{decoy}) < 0;
\label{eqn:ineq}
\end{equation}

\noindent and it is possible to find one such inequality for each
decoy. Expression (\ref{eqn:ineq}) can be rewritten as

\begin{eqnarray}
\sum_{i>j=1}^{20} (n^d_{ij} - n^n_{ij}) \epsilon(i,j) + \epsilon_r
(n^d_r-n^n_r) & \equiv & \sum_{i>j=1}^{20} a_{ij}(\Gamma_{decoy}) \cdot
\epsilon(i,j) + a_r \cdot \epsilon_r\\ & \equiv & {\cal Q}(\Gamma_{decoy},\epsilon)
> 0 \ .
\end{eqnarray}

\noindent where $n^{n,d}_{ij}$ denotes the number of native/decoy
contacts involving amino acids of types $i$ and $j$ and $n^{n,d}_r$
denotes the strength of the native/decoy repulsive term. Given the
native state $\Gamma$ and the sequence $S$, the 211 entries of
$a_{i,j}(\Gamma_{decoy})$ plus $a_r$ depend only on the geometrical
properties of the decoy structure.

For a given set of $M$ inequalities to be satisfied simultaneously, it
is convenient to identify the one (denoted with $l$) that, with the
trial potentials is the worst satisfied one:

\begin{equation}
{\cal Q}(\Gamma_l, \epsilon) < {\cal Q}(\Gamma_k, \epsilon)
\ \ \ \ \ k=1, \ldots, M, \ \ k \neq m
\label{eqn:percstab}
\end{equation}

The selection of the conformation $l$ can be done both when ${\cal
Q}(\Gamma_l, \epsilon)$ is negative (not all the inequalities are
satisfied) and when it is positive (all the inequalities are already
satisfied). ${\cal Q}(\Gamma_l, \epsilon)$ is called the stability of
the set of inequalities for a given choice of $\epsilon$.

Once $l$ has been determined, one updates the trial potentials,
$\epsilon{(i,j)}$ (or $\epsilon_r$) by adding a quantity proportional
to $(a_{ij}(\Gamma_l))$ (or $a_r(\Gamma_l$), where the proportionality
constant is chosen to be much smaller than 1. With this new choice of
the potentials, each inequality is re-valuated and the updating cycle
is repeated. This method can be shown to converge to the optimal
solution: the stability ${\cal Q}$ reaches a constant value (optimal
stability) \cite{perceptron}, which can be of either sign. If it is
negative, it means that no set of potentials can be found that
consistently satisfies all inequalities in the set (unlearnable
problem). To speed up the convergence process towards the optimal
potentials, we found it useful to introduce an additional type of
inequality besides (\ref{eqn:ineq}), namely

\begin{equation}
H(S_i, \Gamma_i) < 0\ .
\label{eqn:ineq2}
\end{equation}

\noindent Such an inequality is useful for ensuring that the native
state of the protein is stable against generally open conformations
(with energy approximately zero) while the previously introduced
inequalities required stability against competing decoys. The
inequalities as in Eq. (\ref{eqn:ineq2}) provide stringent limitations
to the parameter space, thus aiding the search for optimal parameters.

Intuitively, if the number of inequalities exceeds the number of
parameters (211 in our case) and if the energy function is too
simplistic it is unlikely that a solution will be found. Of course,
correlations in the inequalities can make the problem unlearnable even
with very few inequalities or learnable with many of them.

\begin{figure}
\centerline{\psfig{figure=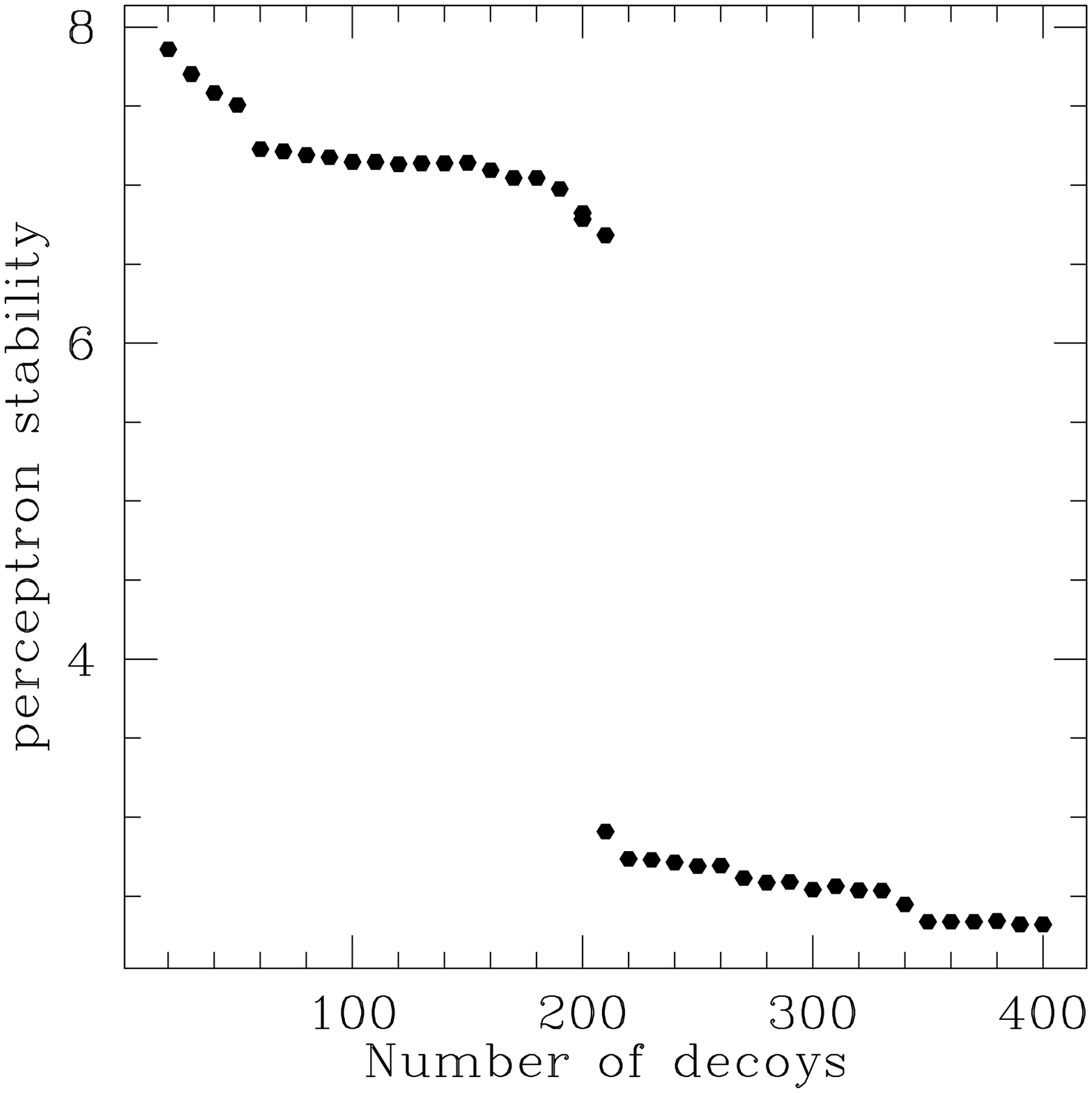,width=3.3in}} 
\caption{Behaviour of the perceptron stability (see
eq. 9) as a function of the number of
decoys generated for protein 1vcc with model A. The first 100
decoys were generated from an initial random set of potentials. Then,
the potentials were refined to provide maximum stability. When the
new set of potentials are used to extract other 100 decoys, a dramatic 
increase in thermodynamic 
stability is observed. The same effect is repeated by iterating the potential
learning
procedure and 
generating other 100 decoys  This proves the effectiveness of updating the
potentials by the appearance of more stringent decoys than the initial ones.}
\label{fig:fig1}
\end{figure}

In any case, the stability threshold cannot increase upon enlarging
the set of decoys/inequalities. As an example we show in
Fig. \ref{fig:fig1} the behaviour of the stability as a function of
the number of decoys obtained through relaxation with the same set of
interaction potentials. It can be seen that due to correlations among
the decoys, the stability does not decrease appreciably after a few
dozens of them have been introduced. The results of
Fig. \ref{fig:fig1} are valuable because they give us an estimate of
the number {\bf $(30-50)$} of representative decoys to be collected at
each iteration step of the potential update.

By updating the potentials, one can generate a set of decoys that are
much closer in root mean square deviation (RMSD) to the target one
than the previous decoys. This is seen in Fig. \ref{fig:fig1} where
the abrupt decrease of stability is visible upon addition of the new
decoys. The iterative potential update can be repeated until the
stability becomes negative or even beyond that. At each iteration
step, it is useful to monitor the average RMSD of the decoys from the
native conformations. It can be anticipated that, as the iterations
proceed, the RMSD will decrease down to a minimal value and then rise
again.  We consider this as a natural cutoff value -- any decoy with a
RMSD below this value ought to be identified with the native state
itself for a given choice of the energy function.  Of course, the
better the choice of the energy functional, the lower this cutoff will
be.

In our study, we have first applied this method to a single protein
for didactic purposes. There after, we considered a set of 20 proteins
(trial set) as representatives of the main folds (see table
\ref{tab:train}) and attempted to learn them simultaneously by storing
100 decoys (5 for each them) before each potential update.

\section{Results and Discussion}

We began by considering a single protein, PDB code: 1vcc, which has 77
residues.  Starting from a set of random potentials, we generated 30
decoys for which we computed the average RMSD, $\bar{g}$, from the
native structure and its variance $\Delta \bar{g}$. With these decoys,
we found the potentials by applying the perceptron algorithm and the
normalization condition ($\sum \epsilon_{i,j}^2 + \epsilon_r^2 = 1$)
which sets the energy scale. With the new interaction parameters, we
generated 30 more decoys and kept repeating this procedure.

\begin{figure}
\centerline{\psfig{figure=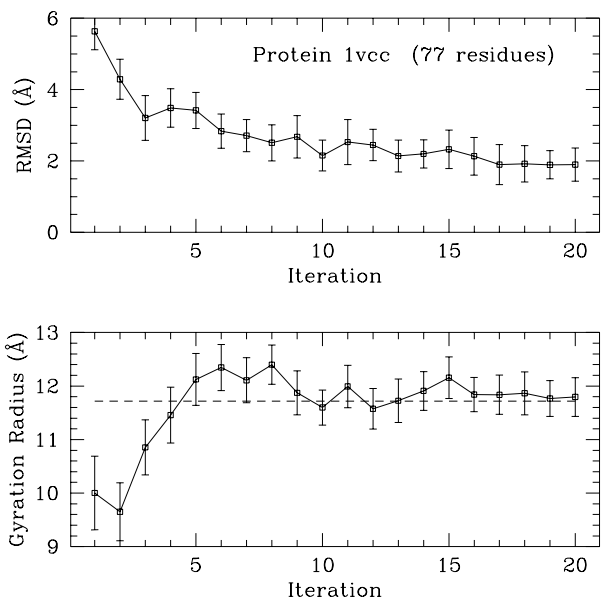,width=3.3in}}
\caption{Results of the iterations on the single protein 1vcc. 
In the upper plot we show the asymptotic RMSD of the decoy structures 
as a function of iteration (potential updates). Each point represents an average over
30 decoys. In the lower plot we show the corresponding radius of
gyration of the asymptotic decoys. The dashed line shows the gyration
radius of the native structure of 1vcc.}
\label{fig:1vcc}
\end{figure}

\begin{figure}
\centerline{\psfig{figure=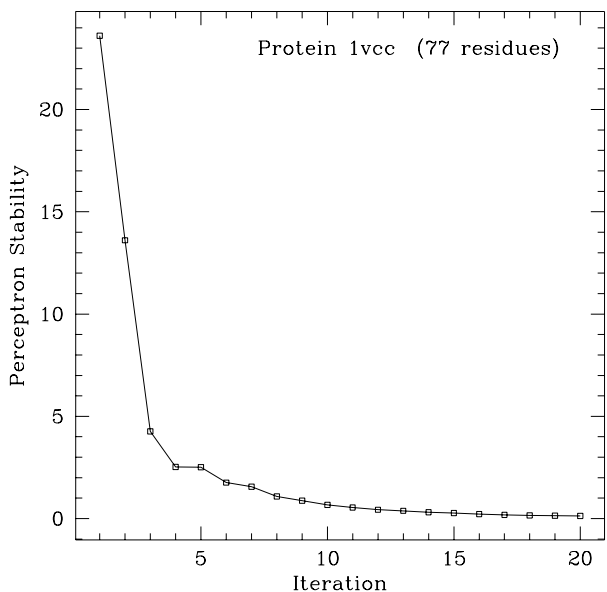,width=3.3in}}
\caption{This plot shows how the perceptron stability decreases as
the iterations proceed on the single protein 1vcc. Extrapolating to an
infinite number of iterations, one obtains negative values for the
stability, consistent with previous observations \cite{Vendruscolo98} that
pairwise potentials are insufficient to stabilize the native state. 
Nevertheless, we show that a single protein can be stabilized within a
narrow native basin (see previous figure).}
\label{fig:stab1vcc}
\end{figure}

In Fig. \ref{fig:1vcc} we show the RMSD, $\bar{g}$, as a function of
the number of iterations. It is remarkable that, with such a
simple model, $\bar{g}$ can be decreased dramatically from the
initial value of about 6 \AA to around 1.5-2.0 \AA, which is just over
the order of the experimental uncertainty! This provides a nice
demonstration of the fact that although, strictly speaking, the
problem of learning the pairwise interactions is unlearnable
\cite{Vendruscolo98} -- for otherwise one would reach zero RMSD - it
is nevertheless possible to stabilize the native state in the native
basin within a low uncertainty. The decrease of perceptron stability
as a function of iteration is shown in Figure \ref{fig:stab1vcc}.

It is instructive to analyze in detail the plot of Fig. \ref{fig:1vcc}
in order to clarify some issues in the use of the perceptron learning
procedure.  At the first iteration, we set the short range repulsive
potential parameter $\epsilon_r$ equal to zero and select random
values for the 210 interaction parameters.  With this choice, very
compact conformations can be reached and consequently the radius of
gyration of the decoys is very small and the RMSD quite high (see Fig.
\ref{fig:1vcc}). However, learning the potentials from this set of
decoys leads to a strictly positive value for $\epsilon_r$ . Indeed,
starting from the second iteration, the radius of gyration approaches
that of the native state and consequently the values of RMSD start to
decrease systematically.  Notably, just a few iterations are
sufficient to set the correct relative scale between the pair
potential interactions and $\epsilon_r$.  This scale, which impacts on
the overall compactness, and is analogous to the average value of the
pairwise interaction cannot be unambiguously determined within other
schemes for potential extraction such as the quasi-chemical
approximation or threading.

Next, we attempted a task considerably more difficult, which was an
attempt to stabilize the native states of twenty proteins
simultaneously. The twenty proteins, shown in Table \ref{tab:train}
were chosen among a list of non-redundant representatives of the main
protein folds. At each iteration step, we generated 5 decoys for each
of them. We saw an improvement as the iterations went on, although not
as pronounced as for the single 1vcc protein with $\bar{g}$ decreasing
to a value of 3.8 $\pm$ 0.5 \AA.  The typical behavior is illustrated
by that of a single protein (chosen to be 1vcc) during the learning
procedure of the potential for the full set of 20 proteins.  (see
Fig. \ref{fig:1vcc20}).  The table of the extracted potential
parameters is given in Appendix A.

As recommended by Lazaridis and Karplus \cite{Karplus2000} in their
recent review article on potential extraction, as an independent test
of the quality of our potentials we assessed their performance on a
set of seven proteins (see first column of Table \ref{tab:rank})
unrelated to those used for extracting the potentials and for which
more than 600 stringent decoys structures (for each protein) have been
derived\cite{Park96}. This unbiased study ought to reflect the
portability of our potentials, i.e. their applicability in contexts
different from which they have been derived.

The test we have performed is the following. For each protein we
compute the energy of the the native state $E_g$ and the energy of all
the decoys $E_i$ ($i=1,\ldots,M$) (where M is the number of decoys for
each single protein) by using our optimal potentials. With the correct
potentials, $E_g$ should be always lower than any other $E_i$. In
table \ref{tab:rank} we report the ranking in energy of the ground
state with respect to all the other decoys. The native state structure
is never the highest ranking one but is always among the best 5 to 10
\%. To better elucidate the quality of our potentials, we have created
a scatter plot (Fig. \ref{fig:evsrmsd}) of the energy of the decoys
(relative to the native state energy) versus the rmsd from the native
state.  From (Fig. \ref{fig:evsrmsd}), the two quantities are seen to
be correlated.  This is a highly non-trivial result since it is
generally difficult to get such correlations even employing specially
designed energy scoring functions \cite{Park96}.

We now turn to a verification of how our potential compares with
others in stabilizing the native state.  To do this we decided to
estimate a new set of potentials by applying the perceptron learning
scheme to the decoys of Park and Levitt. In other words, we identify
the set of potentials which maximally stabilize the ground state of
the seven proteins chosen in ref. \onlinecite{Park96} with respect to
their own competing decoys. With the new potentials we checked the
asymptotic RMSD reached on each of the 20 proteins in our training set
and compared it to the stability obtained with our previous optimal
potentials.  The results are reported in Fig. \ref{fig:confronto} We
repeated the same analysis but working with the seven proteins of the
Levitt data bank and the results are reported in Fig. \ref{fig:sette}.
Remarkably, on the average, the optimal potentials are able to
stabilize the proteins with an accuracy higher than those obtained by
using the potentials extracted using the decoys of
ref. \onlinecite{Park96}.

As another stringent test, we compared the performance of our
potentials with those extracted with the quasi-chemical approximation.
According to the basic prescription of such a scheme, the strength,
$\epsilon_{ij}$, of the interaction between two amino acid types $i$
and $j$ can be deduced from the number of contacting $i$-$j$ pairs,
$f_{ij}$, and the relative abundance of the types, $n_i$ and $n_j$ in
the proteins in the training set. Such potentials are determined up to
a multiplicative constant and an additive one.  We chose the additive
constant of the quasi-chemical potentials by setting their average to
zero. Finally we set the norm of the potential vector to 1, as for the
optimal set. These choices ensure that the two potentials sets are
similar and can be compared on an equal footing. We also checked the
robustness of the extracted potentials by checking that using ten
additional proteins for the extraction procedure did not alter the
potential values appreciably.

\begin{equation}
\epsilon_{ij} = \log {f_{ij} \over (2 - \delta_{ij}) n_i \cdot n_j} 
\label{eqn:quasi}
\end{equation}

Likewise, the strength of the repulsive term, $\epsilon_r$, was
obtained by replacing, in equation (\ref{eqn:quasi}), $f_{ij}$ with
the number of non-consecutive CA pairs that are below 4.5 \AA, while
$n_{i,j}$ is replaced by the total number of CA atoms.
 
We have extracted the quasi-chemical potentials from the twenty
proteins in the training set.  Using this potential, and our
Monte-Carlo procedure, we assessed how well the twenty proteins could
be stabilized.  The results are reported in Fig. \ref{fig:mij}. The
figure shows that quasichemical potentials do not provide the best
stability and their performance is somewhat worse than that obtained
on learning with the Park and Levitt decoys.

As a final test, in order to verify the possibility of improving our
approach, we have introduced a slightly more sophisticated model,
model B, where we consider interactions between all possible pairs of
CA and CB at a sequence separation greater than 1. The specific
interaction is concentrated on the CB atoms, when present. On the
other hand, the interaction between CA atoms is assumed to be
independent of the type of amino acids. Since glycine lacks a CB atom,
a Gly pair will only interact through the CA-CA potentials. For this
reason the CA-CA interaction can be identified in our model with the
Gly-Gly interaction, while interactions between CA-CB will be of type
Gly-X, where X is the amino acid type to which the CB belongs
\cite{Clementi98b}.  In the model, a short range repulsive energy is
now present between all pairs of residues (CA-CA, CA-CB and CB-CB) and
therefore the hard core repulsion described earlier is no longer
needed.

Using this second model, we repeated the same analysis as before and
as demonstrated in Figures \ref{fig:1vccB} and \ref{fig:venti} and in
Table \ref{tab:rank2}, we get an improvement of the results.
This example is helpful in illustrating the possibility of using our
novel optimization technique in selecting the ``most physical'' energy
parameterization.

\section{Conclusions}

We have demonstrated how one may extract effective interaction
potentials between amino acids in a coarse-grained description of a
protein. The method relies on the possibility of finding a set of
competitive decoys of the native state. We outlined an iterative
procedure to generate these decoys which attempts to stabilize, at
least locally, the native state. The results obtained with simple
forms of the energy function are very promising -- we were able to
stabilize a set of 20 proteins to an average distance of less than 4
\AA and moreover, the potentials, when applied to other test with
completely unrelated decoys, yield encouraging results. The use of
slightly more sophisticated and complete forms of the energy function
together with a non-discretized representation of the protein should
lead to even further improvement.

\section{Acknowledgments}

This work was supported by grants from INFM and MURST-COFIN99.

\begin{figure}
\centerline{\psfig{figure=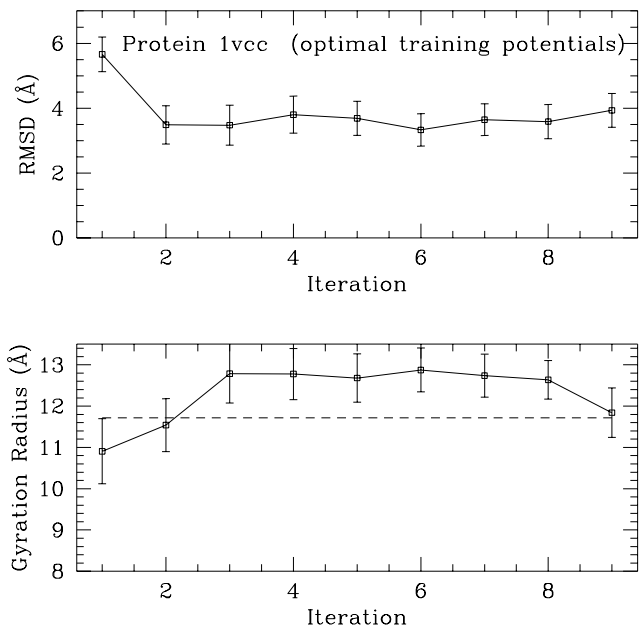,width=3.5in}}
\caption{ Asymptotic RMSD and radius of gyration of the decoy
structures obtained with the protein 1vcc during the iteration
procedure of the ensemble of 20 proteins in Table \ref{tab:train}. At
each iteration stage, 5 decoys are generated and the averages and the
fluctuations are calculated with these 5 decoys.}
\label{fig:1vcc20}
\end{figure}

\begin{figure}
\centerline{\psfig{figure=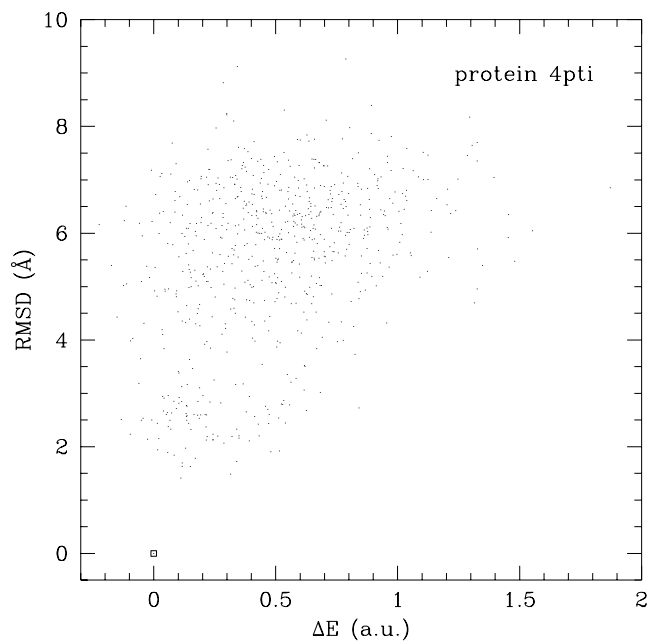,width=3.5in}}
\caption{Scatter plot of the rmsd vs energy gap (in abritrary units)
for the Park and Levitt decoys (there are more than 600 decoys) for
protein 4pti. The energy gap (defined as the difference in energy
between the decoy energy and the true native state energy) was
calculated using the optimal potentials deduced from the stabilization
of the 20 proteins in the training set. Notice that the decoys of Park
and Levitt (and the native state used for comparing the energy) were
not constrained to be on the FCC lattice.}
\label{fig:evsrmsd}
\end{figure}

\begin{figure}
\centerline{\psfig{figure=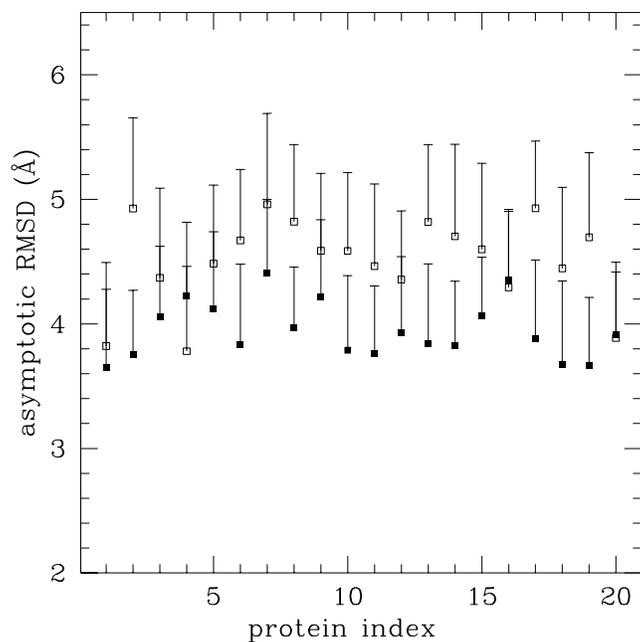,width=3.5in}}
\caption{Average asymptotic RMSD obtained obtained for each of the
twenty proteins of the training set with our potentials (filled
squares) and with those obtained by learning the Park and Levitt decoys (open
squares).  Averages and fluctuations are calculated with 5 decoys
determined by asymptotically relaxing the native state under the
action of the Monte Carlo dynamics. Errorbars are shown only on one
side of the points, to avoid confusing overlaps.}
\label{fig:confronto}
\end{figure}

\begin{figure}
\centerline{\psfig{figure=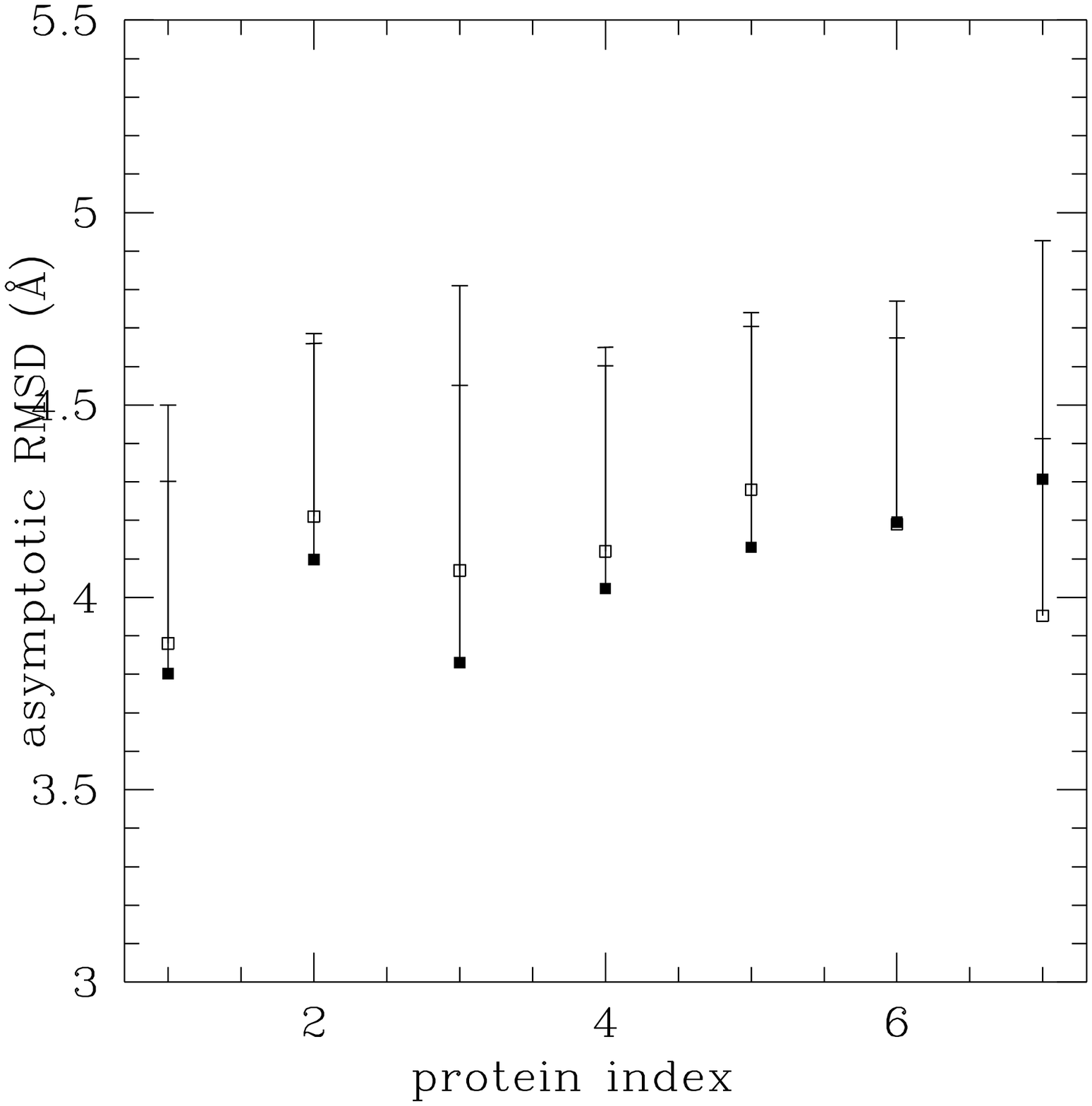,width=3.5in}}
\caption{Average asymptotic RMSD obtained for each of the seven Park
andLevitt proteins with our potentials (filled squares) and with those
obtained by learning the Levitt decoys (open squares) Averages and
fluctuations are calculated with 5 decoys determined by asymptotically
relaxing the native state under the action of the Monte Carlo
dynamics. Errorbars are shown only on one side of the points, to avoid
confusing overlaps.}
\label{fig:sette}
\end{figure}

\begin{figure}
\centerline{\psfig{figure=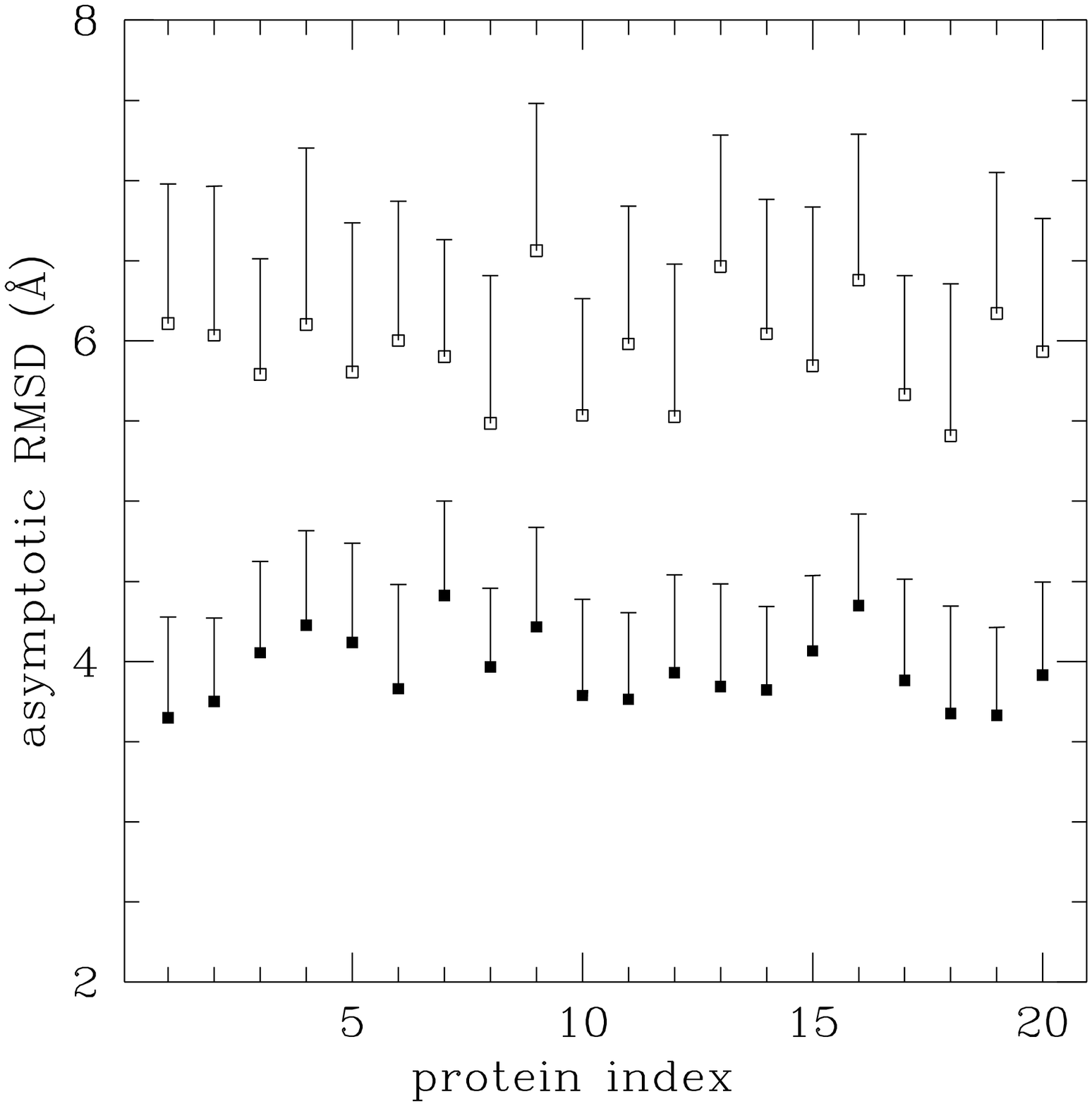,width=3.5in}}
\caption{Average asymptotic RMSD obtained for each of the twenty
proteins of the training set with our potentials (filled squares) and
with those obtained using the quasichemical approach (open squares).
Averages and fluctuations are calculated with 5 decoys determined by
asymptotically relaxing the native state under the action of the Monte
Carlo dynamics. Errorbars are shown only on one side of the points, to
avoid confusing overlaps.}
\label{fig:mij}
\end{figure}

\newpage

\begin{figure}
\centerline{\psfig{figure=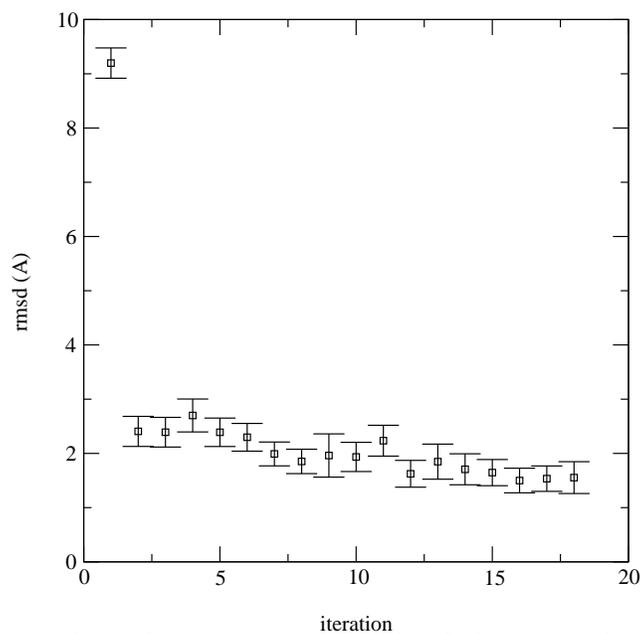,width=3.3in}}
\caption{Results of the iterations on the single protein 1vcc obtained
with the more sophisticated model B described in text.  The asymptotic
RMSD of the decoy structures is shown as a function of iteration
(potential updates).  Each point represents an average over 30
decoys. In this case the value of RMSD is bigger at the first
iteration because, in this model, the $C_{\alpha}-C_{\alpha}$ and
$C_{\alpha}-C_{\beta}$ hard core repulsions are removed and
substituted by a potential that is learned during the iterative
procedure. Initially, this potential is 0 and the resulting
conformations can be extremely compact.}
\label{fig:1vccB}
\end{figure}

\newpage

\begin{figure}
\centerline{\psfig{figure=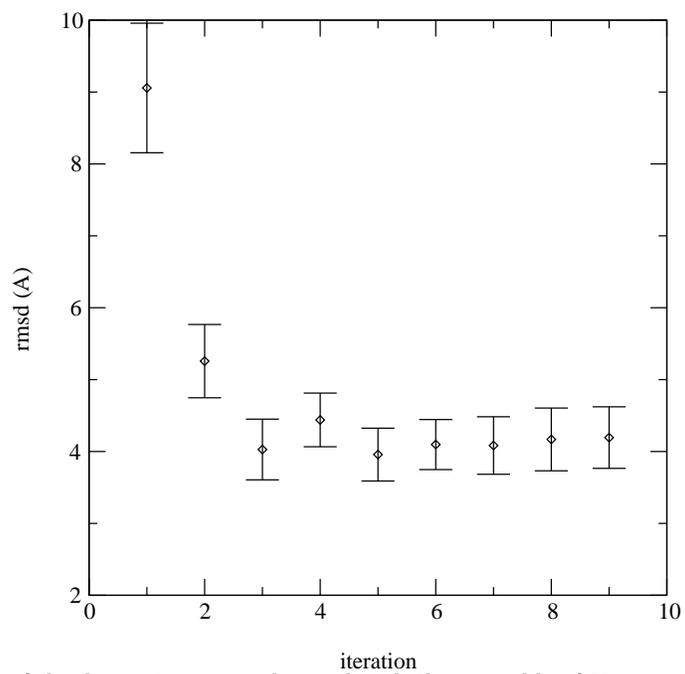,width=3.5in}}
\caption{Asymptotic RMSD of the decoy structures obtained with the
ensemble of 20 proteins in Table I with the model B. At each iteration
stage, 5 decoys per protein are generated and the averages and the
fluctuations are calculated with these 100 decoys.}
\label{fig:venti}
\end{figure}

\newpage

\pagestyle{empty}

\begin{table}[htbp]
\begin{center}
\begin{tabular}{| l | r | }\\
Protein & Length \\ \hline \hline \\
1cbn & 46 \\
1coo  & 81     \\
1erv  & 105     \\
1fna  & 91     \\
1fow  & 76     \\
1hoe  & 74     \\
1hyp  & 75     \\
1kjs  & 74     \\
1mit  & 69     \\
1opd &  85     \\
1pdo  & 129     \\
1rro  & 108     \\
1sap  & 66     \\
1shg  & 57     \\
1tif &  76     \\
1ubi  & 76     \\
1utg &  70     \\
1vcc  & 77     \\
2fxb &  81     \\
2imm &  114 \\
\end{tabular}
\end{center}
\caption{The trial set of 20 single-chain proteins used for extracting
the interaction potentials between amino acids. The set includes
proteins representative of the most common folds and lengths.}
\label{tab:train}
\end{table}

\begin{table}[htbp]
\begin{center}
\begin{tabular}{| l | r | }\\
Protein & Rank\\ \hline \hline \\
1ctf &  31\\
1r69 &   50\\
1sn3 &   12\\
2cro &   31\\
3icb &   75\\
4pti &   26\\
4rxn &   78\\
\end{tabular}
\end{center}
\caption{Ranking (using  model A) of the 7 native proteins for which more than 600
highly competitive decoys have been obtained by Park and Levitt. The ranking is
measured by comparing the native energy (the optimal potentials were
the ones obtained using the 20 proteins in the training set) with those
of the decoys.}
\label{tab:rank}
\end{table}

\begin{table}[htbp]
\begin{center}
\begin{tabular}{| l | r | }\\
Protein & Rank\\ \hline \hline \\
1ctf &  21\\
1r69 &   42\\
1sn3 &   13\\
2cro &   15\\
3icb &   27\\
4pti &   15\\
4rxn &   33\\
\end{tabular}
\end{center}
\caption{Ranking (using  model B) 
of the 7 native proteins for which more than 600
highly-competing decoys have been obtained by Park and Levitt.}
\label{tab:rank2}
\end{table}

\newpage
\tighten

\appendix
\section{Table of optimal interactions}

Below is included the table of the optimal interactions extracted with Model A.
The corresponding value for $\epsilon_r$ is 0.032594 .
\tighten
\begin{tiny}
\begin{table}
\begin{center}
\begin{tabular}{||l|l|r|| ||l|l|r|| ||l|l|r|| ||l|l|r||}
\hline \hline
GLY & GLY &  0.000990 & VAL & ASP &  0.000092 &  MET & PHE &  0.001010 &  HIS & LYS &  0.002934 \\
GLY & ALA & -0.003111 & VAL & ASN & -0.001040 &  MET & PRO & -0.008698 &  HIS & ARG &  0.009985 \\
GLY & VAL &  0.001995 & VAL & GLU &  0.001387 &  MET & TYR & -0.003258 &  HIS & ASP & -0.002501 \\
GLY & LEU & -0.001538 & VAL & GLN &  0.000029 &  MET & HIS &  0.031785 &  HIS & ASN &  0.008099 \\
GLY & ILE &  0.000446 & LEU & LEU & -0.000748 &  MET & TRP &  0.984886 &  HIS & GLU & -0.007232 \\
GLY & CYS &  0.001847 & LEU & ILE & -0.000782 &  MET & SER &  0.002007 &  HIS & GLN &  0.005803 \\
GLY & MET &  0.002339 & LEU & CYS & -0.000196 &  MET & THR & -0.002190 &  TRP & TRP &  0.131813 \\
GLY & PHE &  0.000189 & LEU & MET & -0.002531 &  MET & LYS & -0.004667 &  TRP & SER & -0.002330 \\
GLY & PRO &  0.009071 & LEU & PHE & -0.002127 &  MET & ARG & -0.004676 &  TRP & THR &  0.003848 \\
GLY & TYR & -0.000737 & LEU & PRO & -0.005026 &  MET & ASP &  0.001491 &  TRP & LYS & -0.001668 \\
GLY & HIS & -0.000951 & LEU & TYR &  0.003540 &  MET & ASN &  0.018413 &  TRP & ARG & -0.014845 \\
GLY & TRP & -0.012366 & LEU & HIS & -0.004529 &  MET & GLU &  0.003231 &  TRP & ASP & -0.007832 \\
GLY & SER & -0.003528 & LEU & TRP &  0.010659 &  MET & GLN & -0.002908 &  TRP & ASN & -0.003028 \\
GLY & THR &  0.001084 & LEU & SER &  0.001004 &  PHE & PHE & -0.013128 &  TRP & GLU & -0.009357 \\
GLY & LYS & -0.001308 & LEU & THR &  0.003770 &  PHE & PRO & -0.006986 &  TRP & GLN &  0.012075 \\
GLY & ARG &  0.002466 & LEU & LYS &  0.002119 &  PHE & TYR & -0.003256 &  SER & SER & -0.000802 \\
GLY & ASP &  0.001528 & LEU & ARG & -0.001302 &  PHE & HIS & -0.000190 &  SER & THR & -0.002393 \\
GLY & ASN & -0.001649 & LEU & ASP &  0.000585 &  PHE & TRP &  0.006057 &  SER & LYS &  0.005015 \\
GLY & GLU & -0.000113 & LEU & ASN & -0.000605 &  PHE & SER & -0.001223 &  SER & ARG & -0.001180 \\
GLY & GLN & -0.000425 & LEU & GLU & -0.000453 &  PHE & THR &  0.004102 &  SER & ASP & -0.001609 \\
ALA & ALA &  0.001461 & LEU & GLN & -0.004168 &  PHE & LYS & -0.004479 &  SER & ASN &  0.006249 \\
ALA & VAL &  0.003642 & ILE & ILE &  0.006801 &  PHE & ARG &  0.004855 &  SER & GLU &  0.002888 \\
ALA & LEU &  0.000864 & ILE & CYS &  0.002965 &  PHE & ASP &  0.004899 &  SER & GLN & -0.009002 \\
ALA & ILE & -0.002119 & ILE & MET & -0.009283 &  PHE & ASN &  0.003461 &  THR & THR &  0.003269 \\
ALA & CYS & -0.000751 & ILE & PHE & -0.009792 &  PHE & GLU & -0.001143 &  THR & LYS & -0.005895 \\
ALA & MET & -0.001496 & ILE & PRO &  0.004353 &  PHE & GLN &  0.003790 &  THR & ARG &  0.003967 \\
ALA & PHE &  0.005126 & ILE & TYR & -0.004792 &  PRO & PRO & -0.003621 &  THR & ASP &  0.002193 \\
ALA & PRO & -0.005081 & ILE & HIS & -0.000476 &  PRO & TYR &  0.000996 &  THR & ASN & -0.005914 \\
ALA & TYR & -0.000724 & ILE & TRP &  0.002734 &  PRO & HIS & -0.002032 &  THR & GLU &  0.000948 \\
ALA & HIS & -0.002432 & ILE & SER &  0.001538 &  PRO & TRP &  0.013914 &  THR & GLN &  0.001006 \\
ALA & TRP & -0.009737 & ILE & THR & -0.004179 &  PRO & SER & -0.003125 &  LYS & LYS &  0.005109 \\
ALA & SER & -0.001515 & ILE & LYS &  0.000855 &  PRO & THR &  0.005402 &  LYS & ARG &  0.007273 \\
ALA & THR & -0.000218 & ILE & ARG &  0.001034 &  PRO & LYS &  0.009888 &  LYS & ASP & -0.000642 \\
ALA & LYS &  0.001754 & ILE & ASP &  0.002659 &  PRO & ARG & -0.000067 &  LYS & ASN &  0.006158 \\
ALA & ARG & -0.002511 & ILE & ASN &  0.002317 &  PRO & ASP &  0.000755 &  LYS & GLU & -0.009604 \\
ALA & ASP & -0.001348 & ILE & GLU &  0.007647 &  PRO & ASN &  0.003707 &  LYS & GLN &  0.002349 \\
ALA & ASN &  0.003323 & ILE & GLN & -0.001875 &  PRO & GLU &  0.005402 &  ARG & ARG &  0.009875 \\
ALA & GLU & -0.002376 & CYS & CYS & -0.002544 &  PRO & GLN &  0.000525 &  ARG & ASP &  0.001974 \\
ALA & GLN &  0.005029 & CYS & MET &  0.014331 &  TYR & TYR & -0.007699 &  ARG & ASN & -0.006728 \\
VAL & VAL &  0.001445 & CYS & PHE & -0.013925 &  TYR & HIS &  0.007276 &  ARG & GLU & -0.004586 \\
VAL & LEU & -0.001940 & CYS & PRO & -0.001720 &  TYR & TRP &  0.003708 &  ARG & GLN & -0.001210 \\
VAL & ILE &  0.002618 & CYS & TYR &  0.002585 &  TYR & SER & -0.001895 &  ASP & ASP & -0.000531 \\
VAL & CYS &  0.000296 & CYS & HIS &  0.054553 &  TYR & THR & -0.001235 &  ASP & ASN &  0.007855 \\
VAL & MET & -0.005331 & CYS & TRP & -0.035239 &  TYR & LYS &  0.007956 &  ASP & GLU &  0.002194 \\
VAL & PHE &  0.000008 & CYS & SER &  0.001837 &  TYR & ARG &  0.004237 &  ASP & GLN & -0.001466 \\
VAL & PRO &  0.001362 & CYS & THR &  0.002620 &  TYR & ASP &  0.000182 &  ASN & ASN & -0.001962 \\
VAL & TYR & -0.002175 & CYS & LYS & -0.006040 &  TYR & ASN & -0.006968 &  ASN & GLU & -0.003154 \\
VAL & HIS & -0.006893 & CYS & ARG & -0.006062 &  TYR & GLU &  0.003261 &  ASN & GLN &  0.004502 \\
VAL & TRP & -0.001516 & CYS & ASP &  0.002278 &  TYR & GLN & -0.005137 &  GLU & GLU &  0.006456 \\
VAL & SER & -0.000443 & CYS & ASN &  0.006139 &  HIS & HIS &  0.001314 &  GLU & GLN & -0.005234 \\
VAL & THR &  0.004075 & CYS & GLU &  0.002791 &  HIS & TRP & -0.006739 &  GLN & GLN &  0.008438 \\
VAL & LYS & -0.006987 & CYS & GLN &  0.001387 &  HIS & SER &  0.009858 &  & \\                    
VAL & ARG & -0.005168 & MET & MET &  0.031655 &  HIS & THR & -0.005871 &  & \\                    
\hline \hline
\end{tabular}
\end{center}
\end{table}
\end{tiny}

\end{document}